\documentstyle[aps,preprint,psfig]{revtex}
\begin{document}
\title{\bf Measurement of Isothermal Pressure of Lattice Gas By Random Walk}
\author{Daniel C. Hong$^{}
$\thanks{E-mail address: dh09@lehigh.edu}
and Kevin Mc Gouldrick}
\address{ Department of Physics, Lewis Laboratory,
Lehigh University, Bethlehem, PA 18015 \\}
\maketitle
\begin{abstract}
We present a computational random walk method of measuring 
the isothermal
pressure of the lattice gas with and without the excluded
volume interaction.  The method is based on the
discretization of the exact
thermodynamic relation for the pressure.
The simulation results are in excellent
agreement with the theoretical predictions.
\end{abstract}
\vskip 1.0 true cm
\noindent {\bf I. Introduction}
\vskip 0.2 true cm
The pressure is defined as the force per unit area and is one of the 
measurable macroscopic thermodynamic quantities.  
It is a macroscopic manifestation of the microscopic phenomena of atomic
collisions, by which atoms or molecules transfer the
momentum and thus exert the force to the wall.   
At the macroscopic level,
the pressure can be directly measured by monitoring or
recording the force on the wall of a container.  At the microscopic level, 
a conventional way is to design a method that keeps 
track of the momentum transfered by atoms to the wall
via collisions, and this may be done by Molecular Dynamics(MD)
simulations.$^1$
However, developing and learning MD simulations is time consuming, and
is often well beyond the scope of undergraduate or introductory graduate level
statistical mechanics.  To the best of the authors' knowledge,
no conventional introductory statistical mechanics text books have yet 
dealt with the computational methods of {\it measuring} the pressure.
Considering the surge of computer-assisted course developments in recent 
years, it may be desirable to design a computational method of measuring the
pressure.  The purpose of this paper is to introduce such a
Monte Carlo method, which is
conceptually simple and extremely easy to implement
with a minimal knowledge of comutation,
because it only deals with the random walk
of point particles on a lattice.  We first explain the mathematical basis of
the method in section II and III, and then present simulation results 
for the lattice gas with and without excluded volume interaction 
in section IV.
\vskip 0.2 true cm
\noindent {\bf II. Background}
\vskip 0.2 true cm
Our statring point is the mathematical expression of the pressure P of
a container of volume V at
a given temperature T, which 
can be found in any standard text book of statistical mechanics$^2$,
$$ P = -\partial F/\partial V = kT{\frac{\partial lnZ}{\partial V}}\eqno (1)$$
where $F=-kTlnZ$ is the free energy of the system with Z the partition 
function and k the Boltzmann constant.
The contribution to the partition function comes from two
sources, one from the kinetic energy and the other from 
the potential energy.  The latter 
is known as the configurational integral, which invovles the integration of
interaction energy among particles.  The
former is a function of temperature, and is independent of the volume.
Therefore, at equilibrium,
the contribution coming from the kinetics
is decoupled,
and is separated out from the configurational statistics, and 
the equilibrium pressure can be determined by considering only the
configurational properties of the system.  
The nature of interactions among particles is determiend by
the form of the potential energy.
In this paper, we consider the interaction potential U to be either
nonexistent, in which case the pressure is given by the ideal gas law,
or the particles
interact with each other via hard sphere potential.  Even with hard sphere
potential, the exact expression of the pressure for the
continuum system is not yet known, even though
a few approximate closed forms for the pressure have been reported
in the literature. $^{3,4,5}$
However, for the lattice gas, the situation 
becomes much simpler, because in this case, the free energy can
be computed exactly.
In the lattice version, the hard sphere interaction may be equivalent to
introducing the excluded volume interaction by prohibiting the
multiple occupancy of particles on a given lattice site.  Without such
restriction, the system will follow the ideal gas statistics.  Note that
in both cases, the contribution to the free energy comes only from entropic
part because the interaction energy is zero.  Hence, $ F = - TS = -kT lnZ$,
with S the entropy of the system.
We now compute the entropy of the lattice gas, the result of which will
then be used to test the validity of the random walk method.

Consider now a lattice model, 
where N particles are confined to move only on
d dimensional discrete lattice points of a container with the volume 
$V=H\bullet L^{d-1}$
with H the height and $L^{d-1}$ the area of the side wall, where the
pressure is being monitored.  Note that we set the lattice spacing to the
unit value.
In the absence of excluded
volume interactions, the entropy of the system is given by:

$$ S = kln\Omega = kln(\omega^N) \eqno (2)$$
where $kln\omega$ is the entropy of a single particle, which is simply
the volume V.
Hence, we obtain the ideal gas law. 
$$ P= -T({\frac{\partial S}{\partial V}}) = k{\frac{N}{V}} T = k\phi T 
\eqno (3)$$
with $\phi=N/V$ the density of the particle.
\vskip 0.2 true cm
In the presence of the excluded volume interaction, the entropy of the
system is no longer the product of one particle entropy.  Rather, it
is the total number of ways of putting N particles in a volume V.  In the
discrete version, it becomes,

$$ S = kln\Omega(V,N) \approx -V[\phi ln\phi + (1-\phi)ln(1-\phi)] \eqno (4)$$
where $\Omega(V,N) = V!/N!(V-N)!$
and the Sterling's formula has been invoked.  
The equilibrium pressure is then given by:

$$ P = -T{\frac{\partial S}{\partial V}} = -kTln(1-\phi) \eqno (5)$$
One may use 
the grand partition function approach to obtain the similar result$^2$. 
In this
case, the grand partition function, Q,  becomes:
$ Q=\sum_{N=0}^{\infty}z^NZ_N = (1+z)^V$, where the fugacity
$z=exp(\beta\mu)$ and the canonical partition function
$Z_N=V!/N!(V-N)!$.  Now, from the relations (see Chapter 7 of Huang in ref.2) 
$PV/kT=lnQ$, and $\bar N = z\partial{logQ}{/\partial
z} = zV/(1+z)$, and $\phi = \bar N/V$, we can easily obtain eq.(2).
Note that the pressure of the lattice gas exhibits the logarithmic singularity
at the closed packed density.  For the continuum system, it has been proved 
that the pressure exhibits the algebraic singularity.$^6$
\vskip 0.2 true cm 
\noindent {\bf 
III. The Mathematical Basis for a Random Walk Monte Carlo Technique}
\vskip 0.2 true cm
We now describe a mathematical basis for a Monte Carlo technique that
computes the pressure of the lattice gas numerically.  If the method is
correct, then it should reproduce Eqs. (3) and (5).  This is based on 
a method derived first for the continuum system by Percus$^{7}$,
and then extended to a lattice system
to measure the pressure in polymeric fluids simulations.$^8$
For the sake of completeness,
we will briefly describe the essence of the formalism here, but the reader
is referred to ref.7 and especially ref. 8 where the essential idea has been
advanced.
For a d dimensional
lattice gas in a container of a height H along the z direction
and the wall area $L^{d-1}$, the
derivative becomes the difference equation, and thus the discrete expression
of the pressure is given by:
$$ P = -{\frac{\partial F}{\partial V}} = k(T/L^{d-1})ln(Z(H)/Z(H-1))
\eqno (6)$$
where $Z(H)$ is the partition function with the height H and $Z(H-1)$ is that
with H-1.  
Note that $Z(H-1)$ is
equivalent to $Z(H)$ with
an infinite repulsive potential
at the wall, in which case the
particles cannot appoarch the wall at z=H and thus with U the 
interaction potential between the wall and particles,
the original partition function, $Z(H, U=\infty)$, will reduce to 
$Z(H-1,U=0).$  In order
to utilize this observation in designing the Monte Carlo method,
we
introduce a parameter $\lambda = exp(-U/kT)$
with k the Boltzmann constant.  Note that $\lambda=0$
corresponds to $U=\infty$, and $\lambda=1$ to $U=0$.
Let the total number of particles in the system be
$N$, and $N_w$ be the particles at the wall.  The partition function, $\bar Z$,
for this modified system becomes:
                                                                              
$$\bar 
Z(H,\lambda) = \sum_{E}exp(-E/kT)=
\sum_{N_w =0} ^{L^{(d-1)}} \lambda^{N_w} \Omega(N_w,N-N_w)
\eqno (7)$$
where the summation is taken over 
all possible energy configurations and
$\Omega$ is the total number of configurations
with $N_w$ particles at the wall and 
$N-N_w$ particles
in the bulk.  The probablity, $P(N_w)$,
of finding $N_w$ particles at the wall in the presence
of the potential U is, $P(N_w) = \lambda^{N_w}\Omega(N_w,N-N_w)/\bar Z$.
Further, it is obvious that
$Z(H)=\bar Z(H,\lambda=1)$  and $Z(H-1)=\bar Z(H,\lambda=0)$.
Therefore, if the
mean number of particles at the wall is denoted by
$<N_w>$, then it is given by:
$$<N_w> = \lambda {\frac{\partial ln\bar Z}{\partial \lambda}} \eqno (8)$$
Next, we use a simple mathematical identity to express
the ratio $ln(Z(H)/Z(H-1))$ in terms of $<N_w>$ as:
$$ ln(Z(H)/Z(H-1)) = ln(\bar Z(H,\lambda=1)/\bar Z(H,\lambda=0)) =
\int_o^1 d\lambda {\frac{\partial ln \bar Z(H,\lambda)}
{\partial \lambda}} \eqno (9)$$
Hence, we obtain the desired expression for
the pressure$^{8}$:

$$ P/kT = \int_0^1 {\frac{d\lambda}{\lambda}}\rho_w \eqno (10)$$
where $\rho_w\equiv <N_w>/L^{d-1}$ is the density at the wall. 

\vskip 0.3 true cm
We now demonstrate here that Eq.(10) is
indeed identical to Eqs.(3) and (5).  Consider first the case where
the excluded volume interactions are present.  Since the
partition function $\bar Z$ (Eq.(7)) contains only the configurational 
term, it is
straightforward to write down the expression for $\bar Z$ for
particles in a volume $V=HA \equiv H\bullet L^{(d-1)}$ with $A \equiv L^{d-1}$
the area of the container.  We first consider the
case with excluded volume.  The partition function becomes,
$$\bar Z(H,\lambda) = \sum_{N_w =0} ^{L^{(d-1)}} \lambda^{N_w} \Omega(A,N_w)
\Omega(B,N-N_w) \eqno (11)$$
where $B \equiv (H-1)L^{(d-1)}$.  The first term,$\lambda^{N_w}$ in (A-1) 
is due to the wall potential, and the second and the third terms
are the total number of ways of
putting $N_w$ particles in the wall of area A and $N-N_w$ particles in the bulk
of a volume $B$.
Both $\Omega 's$ in the summation are given by 
the binomial coefficient, i.e.namely $\Omega(Q,R) = Q!/R!(Q-R)!$.       
Let the density of the particle, $N/B = \phi < 1$.  Then, in the
thermodynamic limit, $N_w<<N$.  Using the Sterling's
formula, we find:

$$ ln\Omega(B, N-N_w) \approx -B[ \phi ln \phi + (1-\phi) ln (1-\phi)]
                  + N_w ln[\phi/(1-\phi)] + O(N_w^2/B)$$
Hence,
$$ \Omega(B,N-N_w) \approx exp(Bs_o) [\phi/(1-\phi)]^{N_w} \eqno (12)$$
                                                                               
where $s_o$ is the bulk entropy per site; i.e.
$s_o = -[\phi ln \phi + (1-\phi)ln(1-\phi)]$.               
By putting $(12)$ into $(11)$,
we now obtain the
closed expression for the partition function $\bar Z$:
                                                                 
$$\bar Z = exp(Bs_o) [1 + \lambda \phi/(1-\phi)]^A $$             
        
from which we obtain the exact formula for the density at the wall, $\rho_w$:
$$\rho_w = (\lambda/A){\frac{\partial ln\bar Z}{\partial \lambda}} =
\lambda \alpha(\phi)/(1+\lambda\alpha(\phi))\eqno (13)$$
with $\alpha(\phi) = \phi/(1-\phi)$.  Note that $\rho_w(\lambda=1) = \phi$
as expected.  Hence, we find

$$P/kT = \int _0 ^1 d \lambda \rho_w/\lambda = -ln(1-\phi) \eqno (14)$$

In the case when the excluded volume interaction is not present, the particion
function becomes much simpler:

$$ \bar Z = \sum_{N_w=0}^{N}\lambda^{N_w}\Omega(N,N_w)
A^{N_w} B^{N-N_w} = B^N(1+\lambda A/B)^N \eqno (15)$$
Hence, in the thermodynamic limit of
$A/B<<1$, the density at the wall, $\rho_w$, becomes:

$$ \rho_w = (\lambda/A) {\frac{\partial ln\bar Z}{\partial \lambda}} = 
{\frac{\lambda \phi}{1+\lambda A/B}} \rightarrow \lambda \phi \eqno (16)$$

from which follows the ideal gas law:

$$ P/kT = \int_o^1 {\frac{d\lambda}{\lambda}}\rho_w = \phi \eqno (17)$$

\noindent {\bf IV. Random Walk Simulations}
\vskip 0.2 true cm
We now carry out Monte Calro random walk simulations
to measure the density at the wall and compare them with the
formulas (13) and (16).
\vskip 0.2 true cm
We take a two dimensional lattice of size $H=20$ and $L=10$ with the volume
$V=HL$.  We now
turn on the repulsive wall
potential $0 \le U \le \infty$ at z=H, and introduce a parameter 
$0 \le \lambda = exp(-U/T) \le 1$.  The wall potential is short range and thus
acts only when the particle is right next to the wall, i.e., at $z=H-1$.
At $z=H-1$, the particle moves to the wall at
$z=H$ with the probability $p=\lambda$.  This is the standard metropolitan
algorithm.  Otherwise, the particles are free to move in the bulk
(with the exception of the excluded volume interaction). We also impose
periodic boundary conditions
along the vertical axis.  With these simple rules, 
the Monte Carlo simulations proceed as follows.
Initially N particles with the bulk density $\phi=N/V$ 
are randomly placed on a 
lattice and they undergo random walk.  With the excluded volume interaction,
multiple occupancy on a lattice point is prohibited.  Without the excluded
volume interaction, multiple occupancy is allowed.  When a particle is
right next to the right wall, then it moves there with the
probability $\lambda$.  When an attempt has been made to move all the
N particles once, then one Monte Carlo(MC) step is elapsed.  At each MC
step, the number of particles at $z=H-1$ is registered.  Then,
after M MC time steps, the density of the particles near the wall, $\rho_w$ at
$z=H-1$ is computed as the average over MC time t and configurations, namely:
$\rho_w=  [\sum_{t} N_w(t)/M]/L$ with $N_w(t)$ the muber of particles
at $z=H-1$ at a given MC time t.  $\rho_w$ is then measured
as a function of $\lambda$ for different bulk
density $\phi 's$.  We now present simulation
data.
\vskip 0.2 true cm
We measured 
the particle density at the wall, $\rho_w(\lambda,\phi)$, with and
without
excluded interactions,
for the bulk density $\phi=0.1,0.3,0.5,0.7, 0.9$ 
and for $\lambda = 0.1,0.3,0.5,0.7, 0.9,1.0$.  
By symmetry, $\rho_{wall}(1,\phi) =\phi$.  The simulation data
are presented
in Fig.1 for the excluded volume interaction and
in Fig.2 without the
excluded volume interactions. The dotted lines in both Figures are
the predicted formulas, $(13)$ and $(16)$,
and the squares are the simulation data. 
Even for a small size of lattice used in the simulations, the
agreement between simulation data and the predicted
forumla appear to be quite remarkable. The typical error bars are less than
a few percent
and the error bars are expected to vanish in the thermodynamic limit
as the size increases.
The good agreement between the simulations and the prediction,
even for a small size of lattice, is precisely because of the
fact that the ratio
$A/B$, in equation (16)
is essentially the ratio of one dimensional
lattice sites over those of
two dimensions, which is in the presence
case of order $1/20=0.05<<1$. 
Even for a modest size of two dimensional 
lattice, this ratio is small and its contribution is almost insignificant
in higher dimensions.  Further, the logarithmic singularity in the pressure
at the closed packed density
is also well captured in the lattice simulations with the
typical error bars are less than
a few percent.  We now conclude with two comments.

First, while this method appears to produce excellent results, the
method becomes a little problematic when there is an external bias, say
the gravity toward the wall.  
In this case, unless the
bias is small, it will eventually pull all the particles to the wall, giving
the wall density essentially close to one, and largely
independent of $\lambda$.  In this case, it becomes
quite difficult to exploit the relation (10) for diffent $\lambda$'s.  We have 
not come up with a method to overcome this difficulty.  Such a method,
if successful, will
be quite useful because it will
enable one to directly measure the force profile of granular materials in a
container,  which is the subject of current studies [9-12].  

Second, the method, however, is expected to work quite well in the absence of
bias for other interacting
systems.  An example would be the charged or uncharged
lattice gas with
either long or short range interactions such as Coulomb interactions, or 
other Ising type interactions near the critical point.  Such studies 
will require elaborate finite size scaling analysis and extensive large scale 
simulations, which will be reported in future communications.
\vskip 1.0 true cm
\noindent {\bf VI. Acknowledgements}
\vskip 0.2 true cm
This work
was supported by
NSF as a part of Research Experiences for Undergraduate
Students program at Lehigh University.  DCH wishes to express his thanks to
Ronald Dickman for 
helpful discussions as well as for his independently
obtaining eq.(14).
\vfill\eject

\noindent {\bf References}
\vskip 0.3 true cm
\noindent 1. For example, 
M. P. Allen and D. J. Tildesley, 
{\it Computer Simulations of Liquids},
(Ocford Scientific Publications, 1987)
\vskip 0.2 true cm
\noindent 2. For example, F. Reif,
 {\it Fundamentals of Statistical and Thermal Physics,}
(McGraw-Hill, New York, 1965) p.163.  A little advanced one is:
K. Huang, {\it Statistical Mechanics}, (John Wiley and Sons, 1987).
\vskip 0.2 true cm
\noindent 3. J.K. Percus and G.J. Yevick, ``Analysis of Classical
Statistical Mechanics by Means of Collective Coordinates'',
Phys. Rev. {\bf 110}, 527 (1958)
\vskip 0.2 true cm
\noindent 4. N.F. Carnahan and K.E. Sterling, ``Equation of State for Non-attracting Rigid Spheres'',
J. Chem. Phys. {\bf 51}, 6232 (1970) 
\vskip 0.2 true cm
\noindent 5. J. P. Hansen and I.R. McDonalds, 
{\it Theory of simple liquids}, (Academic 
Press, 1986) p. 95.
\vskip 0.2 true cm
\noindent 6. M.E. Fisher, `` Bounds for the Derivative of the
Free Energy and the Pressure of a Hard-Core Systems near Close Packing,''
J. Chem. Phys. {\bf 42}, 3852 (1965)
\vskip 0.2 true cm
\noindent 7. J. K.  Percus, ``Models for Density Variation at
a Fluid Surface'', J. Stat. Phys. {\bf 15}, 423  (1976)
\vskip 0.2 true cm
\noindent 8. R. Dickmann, ``On the Equation of State of Athermal Lattice
Chains: Test of Mean Field and Scaling Theories in Two Dimensions'', 
J. Chem. Phys. 91, 454 (1989); {\it ibid},
{\it Mumerical Methods for Polymer Systems,} (Springer-Verlag, New York 1997)
p.1.
\vskip 0.2 true cm
\noindent 9. C.-h Liu et al, Science {\bf 269}, 513 (1995)
\vskip 0.2 true cm
\noindent 10. S. N. Coppersmith et al, Phys. Rev. E {\bf 53}, 4673 (1996)
\vskip 0.2 true cm
\noindent 11. S. Luding, Phys. Rev. E {\bf 55}, 4720 (1997)
\vskip 0.2 true cm
\noindent 12. D. Mueth, H. Jaeger, and S. R. Nagel, ``Force Distribution In
a Granular Medium'' (preprint)

\vskip 0.2 true cm
\vfill\eject
%

\noindent Figure captions:
\vskip 1.0 true cm
\noindent Fig.~1.
Comparison between the Monte Carlo data and the exact 
formula (13)
for the lattice gas with the excluded volume interaction for given bulk
density $\phi$'s for different $\lambda 's$.  The vertical axis is the particle
density at the wall, $\rho_w$,  and the horizontal axis is
$\lambda$.  From the bottom to top, $\phi =
0.1,0.2,0.3,0.4,0.5,0.6,0.7,0.8,0.9$.
Dotted lines are the
exact formula(13).
\vskip 1.0 true cm
\noindent Fig.~2. Same as in Fig.1 except that multiple occupancy is 
allowed at the lattice sites.
Dotted lines are the
exact formula(16).
\newpage
\thispagestyle{empty}
\centerline{\hbox{
\psfig{figure=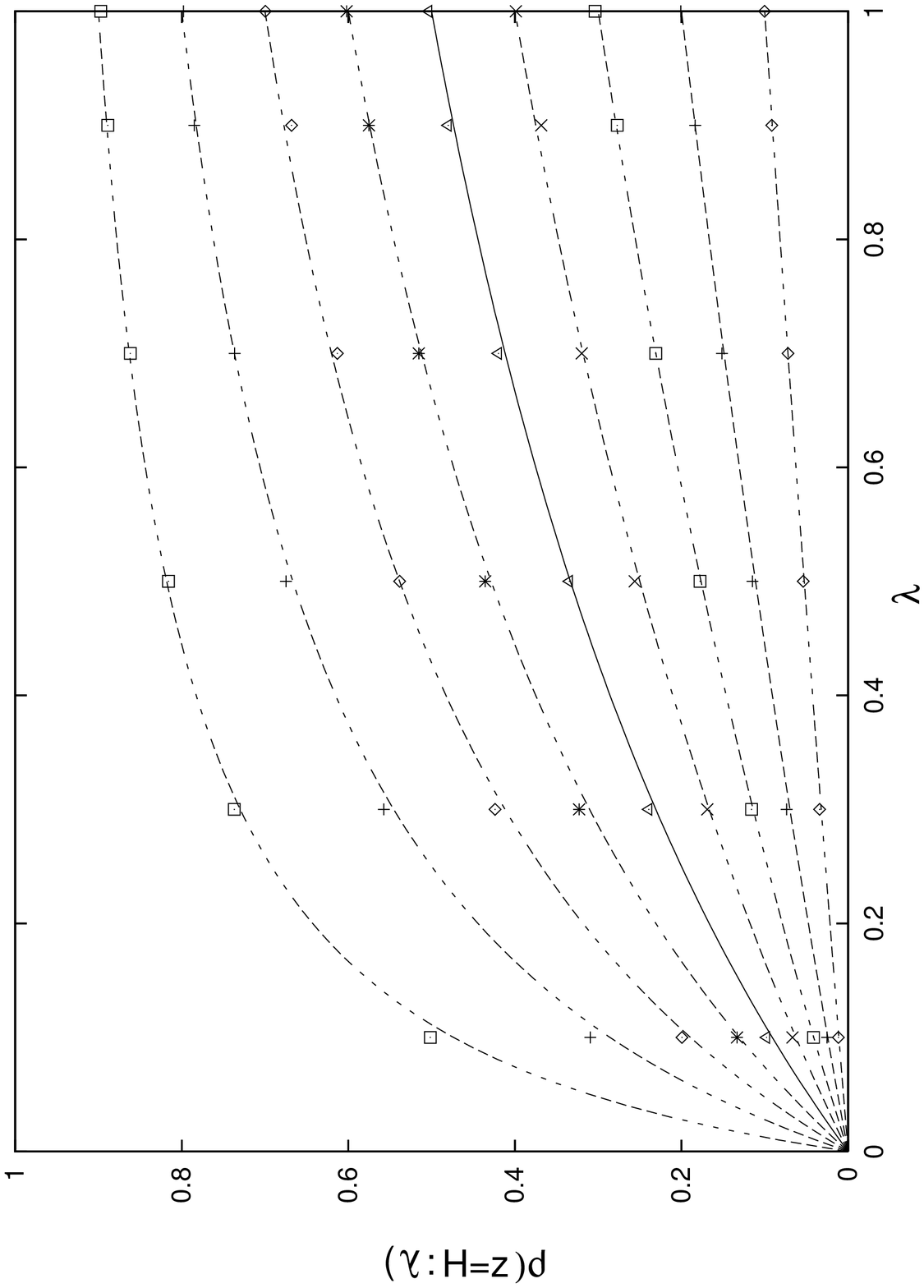 }
}}
\newpage
\thispagestyle{empty}
\centerline{\hbox{
\psfig{figure=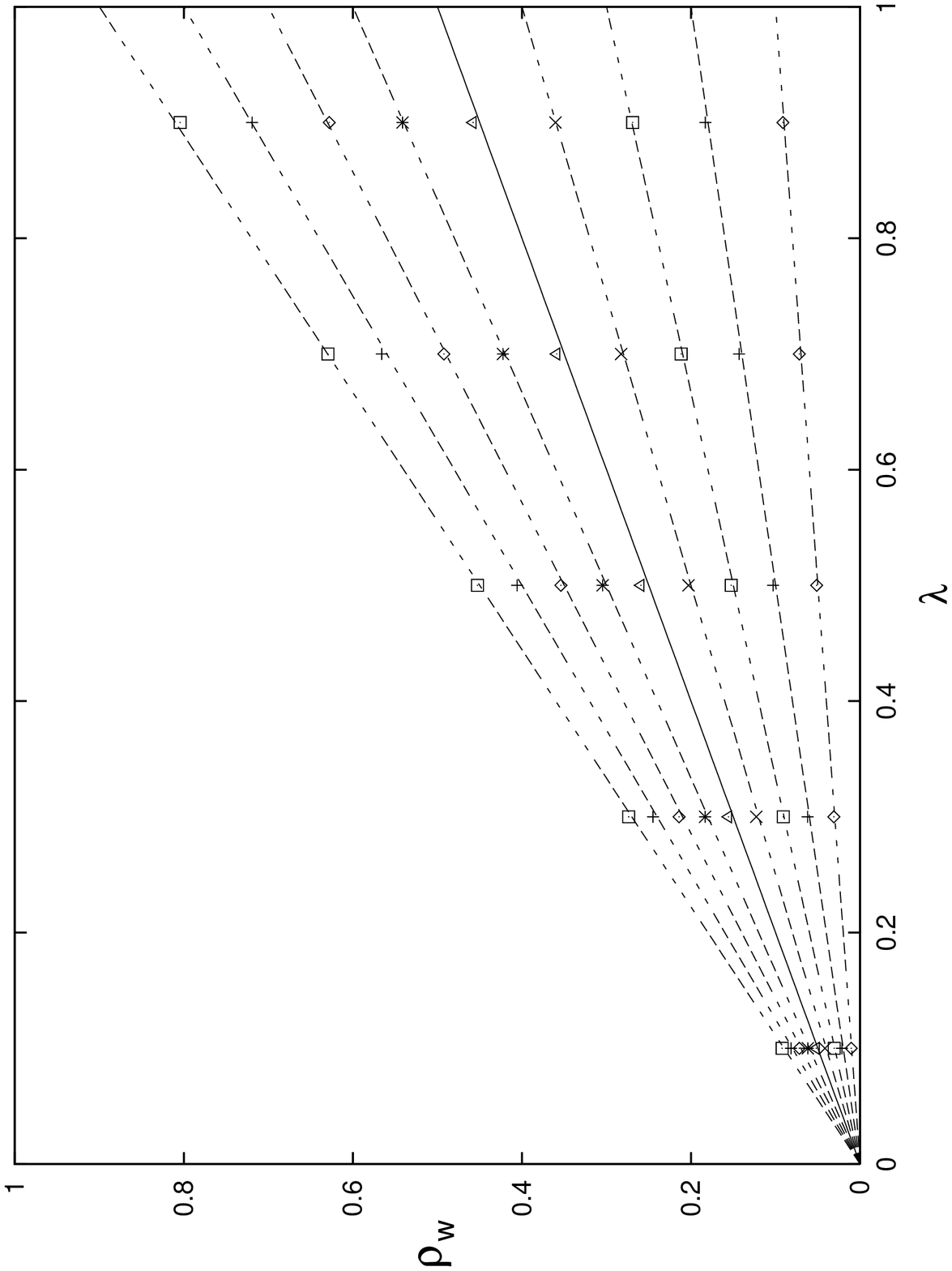  }
}}

\end{document}